\documentclass[journal]{IEEEtran}
\usepackage{hyperref} 

\usepackage{xcolor}
\usepackage{nicefrac}

\usepackage{amssymb}

\usepackage{dcolumn}
\usepackage{bm}
\usepackage[utf8]{inputenc}
\usepackage[T1]{fontenc}
\usepackage{mathptmx}
\usepackage{etoolbox}
\usepackage{cite}

\ifCLASSINFOpdf
   \usepackage[pdftex]{graphicx}
\else
\fi
\usepackage{amsmath}

\usepackage{url}
\usepackage{xcolor}
\usepackage{soul,xcolor}

\begin{document}
%
\title{Detection of low-energy fluxons from engineered long Josephson junctions for efficient computing\\}
%
%
%

\author{ 
        Han~Cai,
        Liuqi~Yu,
        Waltraut~Wustmann,
        Ryan~Clarke,   
        and~Kevin~D.~Osborn 

 
\thanks{H. Cai and L. Yu are with the Quantum Material Center and the Department of Physics at the University of Maryland, College Park, MD 20742, USA }

\thanks{W. Wustmann is with the Laboratory for Physical Sciences, University of Maryland, College Park, MD 20740 USA, and the Department of Physics and the Dodd-Walls Centre for Photonic and Quantum Technologies, University of Otago, Dunedin 9016, New Zealand.}

\thanks{R. Clarke is with the Laboratory for Physical Sciences, University of Maryland, College Park, MD 20740, USA}

\thanks{K. D. Osborn is with the Laboratory for Physical Sciences, University of Maryland, College Park, MD 20740, USA, and the Joint Quantum Institute and Quantum Materials Center, University of Maryland, College Park, MD 20742, USA (email: osborn@lps.umd.edu)}

}

\maketitle
\setstcolor{red}

\begin{abstract}

Single-Flux Quantum (SFQ) digital logic is typically energy efficient and fast, and logic that uses ballistic and reversible principles provides a new platform to improve efficiency. We are studying long Josephson junctions (long JJs), SFQs within them, and an SFQ detector, all intended for future ballistic logic gate experiments. Specifically, we launch low-energy SFQ into engineered long JJs made from an array of 80 JJs and connecting inductors. The component JJs have critical currents of only 7.5 $\mu A$ such that the Josephson penetration depth is approximately 2.4 unit cells, and the SFQ's stationary energy in the LJJ is $\sim$47~zJ. The circuit measured consisted of three components: an SFQ launcher, the LJJ, and an SFQ detector that uses JJ critical currents of only 15-20$\mu A$. The circuit was measured in two environments: at 4.2~K in a helium dunk probe and 3.5~K in a cryogen-free refrigerator. According to calculations, the SFQ may traverse the LJJ ballistically, i.e., with a small change in velocity. Data show that SFQ detection events are synchronous with SFQ launch events in both setups. The jitter extracted from the launch and arrival times is predominantly attributed to the noise in the detector. This study shows that we can create and detect low-energy SFQs made from engineered LJJs, and the importance of jitter studies for future ballistic gate measurements.  

\end{abstract}


\section{INTRODUCTION}

\IEEEPARstart{T}{he} size and number of computations are increasing steadily and, consequently, there is a growing need to study the energy efficiency of computing. According to current predictions, the power consumption of an exascale computer with CMOS circuits will be more than 20~MW \cite{shalf2011exascale}, and require a significant infrastructure for cooling such that various alternative technologies must be investigated. Due to its use of magnetic flux quantization and the different types available, superconducting electronics (SCEs) offer a wide range of alternative computational approaches that contrast industry-standard logic. For years, SCEs have enabled research-grade demonstrations with higher operational speed and lower power consumption than standard CMOS logic \cite{mukhanov2011energy}. Recently, ballistic\cite{wustmann2020reversible, frank2017asynchronous} logic provides insight into computation with a new computing platform.  SCEs generally includes Josephson voltage sources \cite{burroughs2011nist, donnelly20191}, digital receivers \cite{kirichenko2009microwave}, and studies of devices for neuromorphic computing \cite{crotty2010josephson, schneider2022supermind}, network switches \cite{kameda2005high}, astronomical sensor readouts \cite{sahu2019low, bozbey2009single}, and qubit control \cite{leonard2019digital, mcdermott2014accurate}.

The most developed SFQ logic is irreversible \cite{tanaka201218, herr2011ultra, dorojevets2014towards, kirichenko2011zero}, but reversible logic offers the highest energy efficiency in a given system of materials. Physically reversible gates have the potential to break new efficiency records, and they can bring the energy cost per gate operation below ln(2)$k_B T$. This entropy limit is imposed on a logic gate that erases one bit. It is also comparable to the energy scale of thermal fluctuations $k_B T$, which relates to thermally induced errors. 

Previous demonstrations of reversible logic primarily use adiabatic waveforms for power, including the negative-inductance SQUID (nSQUID) \cite{semenov2003negative} and the reversible version of the adiabatic quantum flux parametron \cite{takeuchi2013measurement}. The nSQUID was demonstrated with DC generation of SFQs that propel bits forward \cite{ren2011progress}. In contrast, AQFP is AC powered and normally demonstrated without logical reversibility, but it reportedly realizes an average energy cost of 1.4~zJ per gate operation \cite{ayala2020mana} at 4.2~K. Furthermore, a reversible AQFP cell operation has been demonstrated \cite{takeuchi2015thermodynamic}, and simulations show that this can reach below Landauer's limit in energy cost if the clock speed is sufficiently lowered. \cite{takeuchi2013simulation}. Despite the success of AQFP, the power uses multiple clock phases, which create design complexity compared to DC-powered logic types, including Rapid Single-Flux-Quantum (RSFQ) logic and variants. 
 
As an alternative to adiabatic reversible logic, one may in principle use ballistic bits for reversible SFQ logic\cite{osborn2020reversible, wustmann2020reversible, osborn2023asynchronous, frank2017asynchronous}. This class uses SFQ in long Josephson junctions (LJJs) as bit states into and out of gates. The SFQ may also be called fluxons (or flux solitons) in the LJJs, due to their particle-like properties (within the LJJs). The logic type we study, Reversible-Fluxon Logic (RFL), relies on a novel nonlinear resonance between pairs of LJJs to realize gate operations\cite{wustmann2020reversible}. The most fundamental gate operation with this resonance is the NOT or inversion gate, which reverses the fluxon's polarity and reaches an energy efficiency over 97 $\%$.  Furthermore, from simulations, we developed a CNOT gate \cite{osborn2020reversible} and a class of asynchronous gates named ballistic shift registers \cite{osborn2023asynchronous}. In other work, abstract models of 2- and 3-terminal ballistic gates were studied \cite{frank2017asynchronous, frank2019asynchronous} for universal computing with ballistic particles.

In this study, we designed engineered LJJs as media for low-energy fluxons, measured fluxons transmitted through the LJJs, and analyzed the results. The engineered LJJs contained undamped Josephson junctions (JJs) in an array, where each cell within the array contains one JJ. To test the fluxon transmission through LJJs, we use a fluxon launcher circuit before the LJJ and a fluxon detector circuit after the LJJ. We tested LJJ transmission at 4.2~K in a helium dunk probe (DP) and at 3.5~K in a cryogen-free refrigerator (CFR). Our study indicates that the discrete LJJ design likely limits the loss and that the detector's performance is limited by noise. 

\section{EXPERIMENT AND RESULTS}

For a reliable design of circuits, we have formulated our LJJs using component inductors and JJs in a quasi-lumped design. The LJJ design is chosen to allow integration into RFL gates later. In this design, our Josephson penetration depth extends over two unit cells and thus the core of the fluxon extends over 4 cells, where a unit cell is the periodic unit of the engineered LJJs.  We study the LJJs using two nominally identical chips and the two experimental setups. The two measurement setups provide different noise environments in which to study low-energy fluxons.
 
\subsection{LJJ parameters, component analysis, and launch speed}

\begin{figure*}[htbp]
\centerline{\includegraphics[width=2\columnwidth]{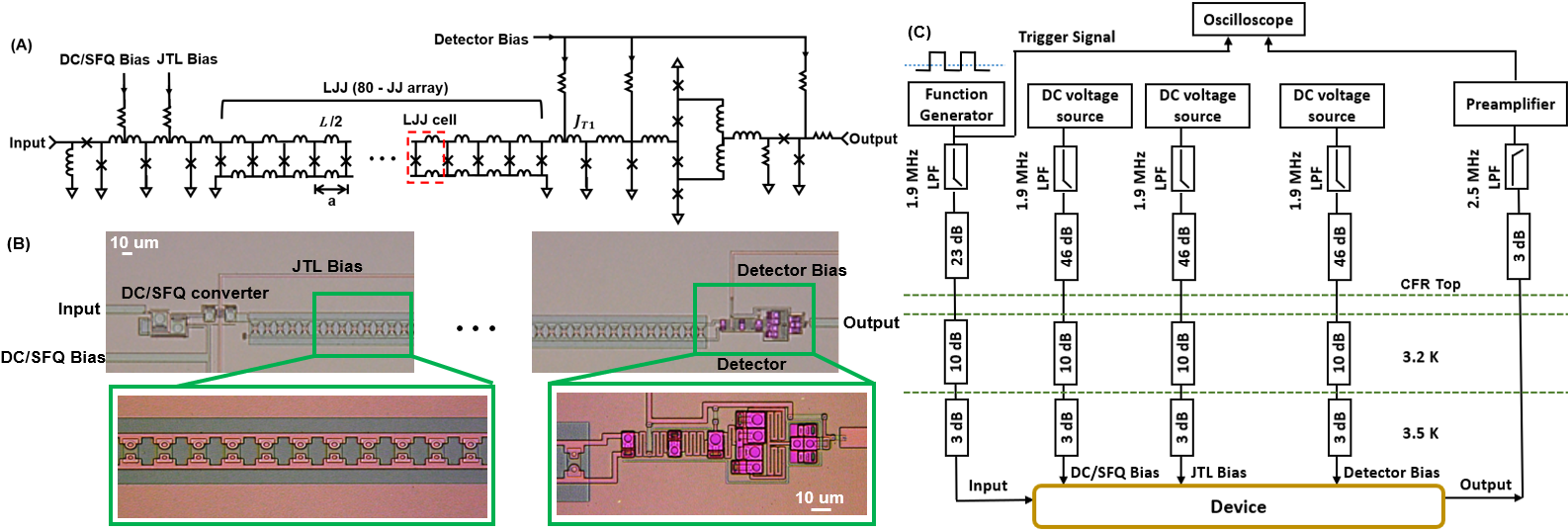}}
\caption{Engineered LJJ test circuit. (A) The circuit diagram shows the DC/SFQ launcher, the LJJ, and the SFQ detector. The LJJ is composed of an array of 80 JJs spaced by length $a$ and connecting rail inductors. Due to the shape of the LJJ, the connecting inductors are named rails, as an analogy to rails of a ladder, and the JJs are in wires are called rungs, as an analogy to rungs of a ladder. The rail segments are inductors between JJs, each with a value of $L/2$ such that the total inductance per cell is $L$. (B) Optical photographs of the fabricated circuit, including enlarged photographs of the discrete LJJ cells and the detector. The JJs in the LJJ have a critical current of 7.5 $\mu$ A. The detector is an SFQ/DC converter and a two-JJ Josephson Transmission Line (JTL) stage at the detector's input. The JJs have a critical current of approximately 20~$\mu$A. The first JJ constitutes a critical part of the detector, since it must forward the fluxon inward to the rest of the detector, and it is subject to thermal and other noise. Along with some bias at the end of the LJJ, the JTL stage is intended to power the fluxon further onward to a T flip-flop in the detector. The detector's JJs are approximately five times smaller than the JJs on the launcher (DC/SFQ converter) to allow the detection of low-energy fluxons. In the optical photographs, the LJJ shows holes in the ground plane underneath the LJJs. (C) A schematic drawing of the CFR measurement setup; the DP setup is comparable (see main text). A 3~dB attenuator is inserted before the preamplifier's high-impedance input to reduce rf-reflections back toward the on-chip detector.}
\label{circuit}
\end{figure*}

Circuits in this work were fabricated utilizing $Nb/AlO_x/Nb$ trilayer Josephson junctions and other elements at a superconducting digital foundry (see acknowledgment).  The main circuit in this work allows us to study fluxons traveling through our LJJs. The circuit consists of a DC-to-SFQ (DC/SFQ) converter that functions as a fluxon launcher, an LJJ, and an SFQ-to-DC (SFQ/DC) converter that functions as a fluxon detector. The detector circuit gives measurable DC levels which change upon arrival of a fluxon (see Fig. \ref{circuit}). A fluxon will be launched into the LJJ when the rising edge of the input waveform arrives at our DC/SFQ converter (launcher). The SFQ/DC converter (detector) uses a T flip-flop (TFF) to alternatively switch between two internal states upon each fluxon arrival \cite{polonsky1993new}. As a result, every edge of an output waveform corresponds to a fluxon arrival, and the detector signal frequency should be half of the launcher input frequency.

Fig.~\ref{circuit} (A) and (B) depict a schematic and some photographs of the test circuits, respectively. The left enlarged view in Fig. \ref{circuit} (B) illustrates discrete LJJ cells consisting of niobium trilayer Josephson junctions (JJs) and interconnecting "rail" inductors. Inductor wires have an extracted inductance of $L/2 = 3.92$~pH per rail per cell, where L is the inductance per cell. There are 79 complete cells in the LJJ, with one JJ per cell (one JJ in the LJJ is not part of a complete cell). The critical current used in this paper is calculated from the designed and nominal current density, rather than the actual current density. The nominal critical current density, $J_c$, is 1~$\mu$A/$\mu \text{m}^2$, and the nominal JJ area per LJJ cell is 7.5 $\mu \text{m}^2$.  Consequently, the nominal critical current of the JJ in LJJs is 7.5~$\mu$A. The unshunted JJ inductance, $L_J$ = $\Phi_0/2\pi I_C$, is 43.88~pH. The ratio of inductance between the JJ inductance and cell inductance $L_J/L$ is 5.64, such that the fluxon can, in principle, travel ballistically, i.e., with a small change in velocity. The Josephson penetration depth is 2.4 unit-cells ($\lambda_J$ = $a\sqrt{L_J/L} $) and the LJJ has 79 unit cells such that the media is long compared to a fluxon.


We determine the rail-segment inductance $L/2$ in our LJJ by measuring the inductance of a wire fabricated on the same chip as our LJJ structures. The measurement is of the SQUID voltage versus direct-injected flux bias current ($I_{fl}$). From this, we determine the self-inductance from $L_s=\Phi_0/\Delta I_{fl}$. The SQUID inductance tested has the same linewidth as the narrowest wire in the rails of our LJJ, which are $2 \mu $m wide. The narrowest segment dominates the inductance and has a length of $a' = 6 \mu \text{m}$, less than the total cell length ($< a \approx 16 \mu \text{m}$). The wide linewidth in the rail inductors is $6 \mu \text{m}$.  For 2~$\mu \text{m}$ linewidth wire, the SQUID-measured inductance per length is 0.654~pH/$\mu \text{m}$, with a standard deviation of 3.8 $\%$ (see Appendix A for details). The extracted inductance of the LJJ rail segments follows from the SQUID-measured inductance per length and the length $a'$.

The fluxon may lose energy to plasma waves as it moves through the LJJ, due to finite discreteness\cite{pfeiffer2008resonances}. However, in our structures, we expect the energy loss and related velocity change to be small \cite{wustmann2020reversible, yu2019experimental}. For example, in previous experimental structures  \cite{yu2019experimental}, a typical discreteness parameter used was $a/\lambda_J = \sqrt{1/7}$. This small ratio of $a/\lambda_J$, provides small (negligible) discreteness for fluxon transmission \cite{wustmann2020reversible}. The ratio found in our LJJs is $\sqrt{1/5.64}$. Simulations of our structure show the fluxon may move ballistically at our experimental LJJ discreteness and launch speed. Using the nominal critical currents and the extracted rail inductance, we simulated the SFQ generation and launch process. Our simulations give a launch velocity of $v \approx 0.78c$, where $c$ is the Swihart velocity \cite{swihart}. This velocity is higher than the velocity used for typical gate simulations, which is $v=0.6c$. Larger speeds give more energy loss to plasma waves from discreteness, but still, our structure is short enough to induce small losses such that the fluxon may propagate ballistically. 


\subsection{Measurement of reference and LJJ-containing devices with the DP setup}

\begin{figure}[htbp]
\centerline{\includegraphics[width=0.8\columnwidth]{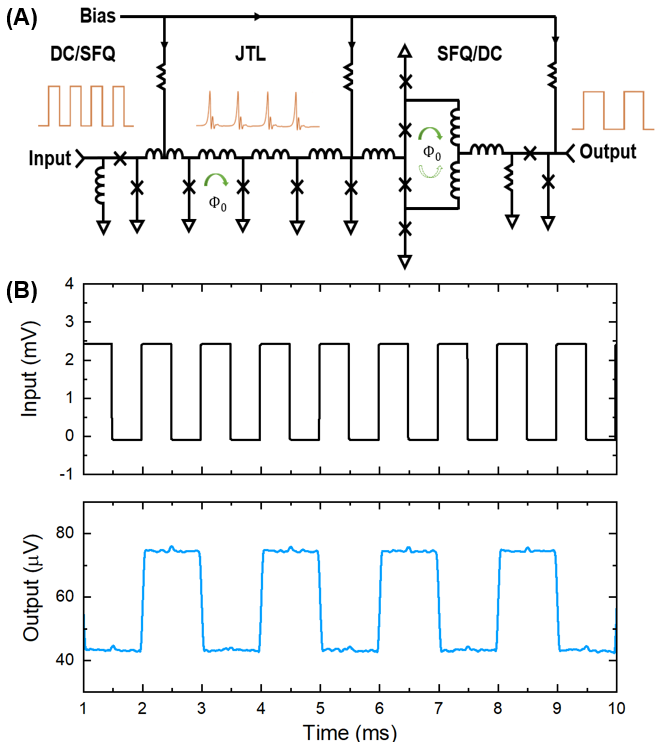}}
\caption{A reference DC/SFQ and SFQ/DC converter circuit without LJJs. (A) Circuit schematic. The circuit includes a DC/SFQ converter that generates SFQ pulses. These pulses are injected into a short JTL and then detected by an SFQ/DC converter. The JJs in this circuit have a critical current of approximately 100~$\mu$A. The bias currents are set at roughly 75$\%$ of the critical current of the Josephson junctions, providing sufficient amplitude to trigger the junctions in response to an incoming SFQ pulse. The central element of the SFQ/DC converter is a T flip-flop, which changes state upon receiving an SFQ pulse. After a second SFQ (and all even-numbered SFQ), the output toggles back to the superconducting (zero-voltage) state. Therefore, two periods of the input waveform are designed to produce one period in the output waveform. (B) Test results from a 1~kHz input waveform (upper panel) show the proper operation of the converter circuitry in our DP setup (lower panel). }
\label{converter}
\end{figure}

Before testing the LJJs, we conducted tests of converters without LJJs. The test circuit for this includes a DC/SFQ converter, which generates an SFQ from every input DC pulse, and an SFQ/DC converter, which detects the arrival of an SFQ pulse. A 100~$\mu$A-JTL section is used to connect these two converters. The chip schematic is shown in Fig.~\ref{converter} (A). In the DP measurement setup, the chip is mounted on an RF dunk probe and initially tested in a screen room to shield the digital circuit from ambient RF noise. However, we did not see a substantial difference in noise level when the screen room door was opened. The DP is immersed in 4.2~K liquid helium, and its full measurement setup contains attenuators at room temperature to reduce noise on the chip, with values of 20~dB on the input signal line and 56~dB on the bias line. This setup is comparable to the CFR setup of Fig.~\ref{circuit} (C), and we explain the difference between the two setups in the discussion section below. We sent a square waveform to the input converter and a DC bias to each converter. On-chip resistors converted bias voltages to currents. We measured the output signal through a 2.5~MHz low-pass filter, followed by a preamplifier with 1-MHz bandwidth. To reduce RF reflections on the output cable, we added a 3~dB attenuator in front of the filter that connects to the preamplifier input. Fig.~\ref{converter} (B) shows the input and output waveforms, where the output waveform is shown after dividing by the preamplifier's gain. The output waveform frequency is half of the input waveform frequency and the waveforms are synchronized. This behavior indicates the creation of an SFQ in the DC/SFQ converter and a corresponding detection event on every input clock cycle.

Afterward, we conducted tests on LJJs in the DP setup. For this, we use the previously described launcher and setup of Fig.~\ref{converter}. The launcher was made to send SFQ through JTL sections with JJ critical currents of 100~$\mu$A. However, the JJs in our LJJ have a nominal critical current $I_C=7.5$~$\mu$A, and the central portion of half of the fluxon's energy is stored within approximately 5 unit cells within the LJJ, resulting in the fluxon's energy being lower than that of an SFQ made from a couple of 100~$\mu$A JTL sections. To detect incoming fluxons, we used JJs with relatively small critical currents. We chose JJs that are five times smaller for our fluxon detection circuit than our launcher. The currents are also smaller than the reference converter circuit of Fig.~\ref{converter}. 

\begin{figure}[htbp]
\centerline{\includegraphics[width=0.8\columnwidth]{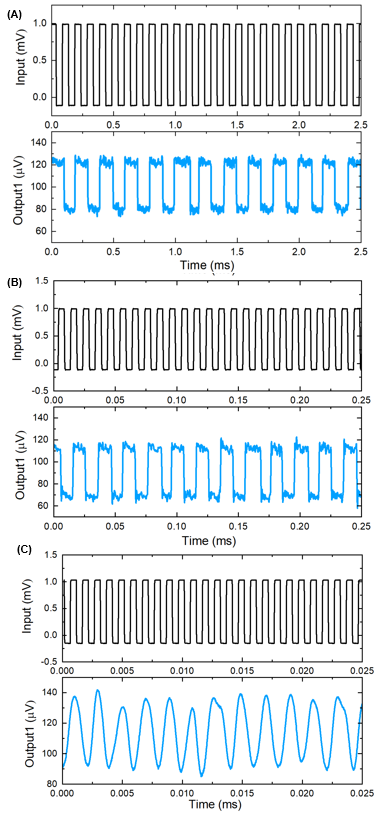}}
\caption{LJJ test-circuit data taken with the DP setup, shown with the corresponding input waveforms. (A), (B), and (C) show the converter output voltage for input fluxon frequencies of 10~kHz, 100~kHz, and 1~MHz, respectively. Similar to the circuit in Fig.~\ref{converter}, the output signal frequency is half of the input signal frequency. At low frequencies, the output amplitude is approximately 50~$\mu$V.}
\label{DPdata}
\end{figure}

We measured the output waveform for the LJJ test circuit and found an amplitude of approximately 50~$\mu$V. Data was obtained up to 1~MHz input frequency, with representative data shown in Fig.~\ref{DPdata}. As the input frequency increases, the preamplifier's 1~MHz maximum bandwidth limits the high harmonics of the output square wave, causing the output to appear similar to a sinusoidal waveform. By counting the number of periods of the output data, we found that the output frequency was half of the input frequency, indicating that fluxons passed through the LJJ. However, this measurement data shows a relatively high variation in the time between each transition (higher jitter). We will discuss the jitter in detail below. Next, we show results from the same chip in the other setup.

\subsection{Measurement of LJJs with the CFR setup}

\begin{figure}[htbp]
\centerline{\includegraphics[width=0.8\columnwidth]{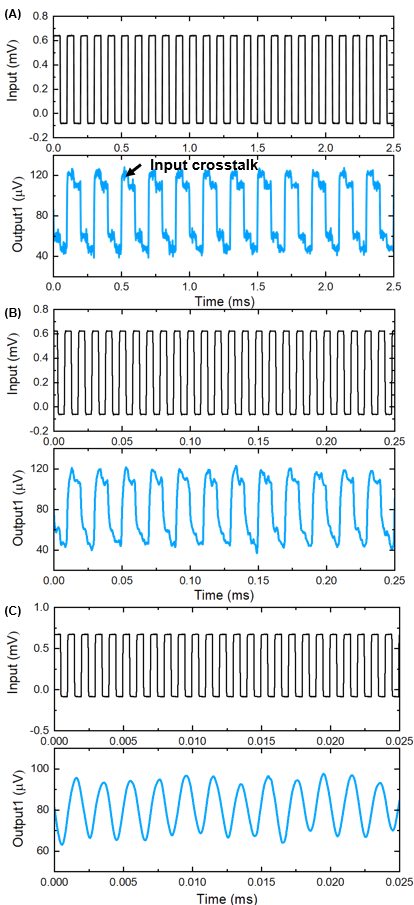}}
\caption{LJJ test-circuit data taken with the CFR setup, shown with the corresponding input waveforms. (A), (B), and (C) show the converter output at operating frequencies of 10~kHz, 100~kHz, and 1~MHz, respectively. Input-to-output crosstalk is observed at the lowest frequency. At 1~MHz, the arrival of the input waveform at the chip and the output waveform from the chip are attenuated by input and output filters. The output-line filtering transforms the 1~MHz square wave nearly into a sine wave. As expected, the output frequency is half of the input signal frequency. The output amplitude at low frequency is approximately 65~$\mu$V.}
\label{CFRdata}
\end{figure}

The same LJJ-test chip was next installed in the CFR, where the CFR measurement setup is shown in Fig.~\ref{circuit} (C). This setup has 13~dB of attenuation at low temperature on all input lines, a 23~dB attenuator at room temperature on the input signal line, and a 46~dB attenuator at room temperature on the bias lines. Additionally, the setup has a 1.9~MHz low-pass filter on input lines to lower noise from room temperature. 

Similar to measurements from the DP setup, we measured the output waveform from the CFR setup (see Fig.~\ref{CFRdata}). Again we find the capability of operating at frequencies up to 1~MHz. The preamplifier's 1~MHz bandwidth smooths the output signal waveform and lowers the amplitude above 1~MHz. The output amplitude is approximately 65~$\mu$V at low frequencies.  The difference in output amplitude between the two setups is due to a different bias value applied to the detector. 

The operation margins of a 10~kHz input signal amplitude, DC/SFQ bias, input JTL bias, and detector bias (see Fig.~\ref{circuit}) are 23.4$\%$, 18.25$\%$, 11.57$\%$, and 8.7$\%$, respectively. Input-to-output crosstalk is observed (shown in Fig.~\ref{CFRdata}). We have not yet determined the cause of the crosstalk, but it may be allowed by the absence of a shorting connection from the sky plane to the ground plane at the detector. We note that our LJJs have no ground plane under them, and this may allow increased crosstalk between different parts of the circuit relative to a circuit with a ground plane. Compared to the results obtained from the DP setup, the data here exhibit a relatively stable pulse width (low jitter), indicating that the CFR provides a lower noise environment for our LJJ circuit than the DP.  


\subsection{Jitter measurement} 


Each rising transition in the input waveform generates one fluxon at the launcher. Additionally, both rising and falling transitions in the output waveform indicate fluxon detection. To calculate the jitter, we subtract times between input and output waveforms corresponding to fluxon generation events on the input waveform and detection events on the measured waveform. The jitter is analyzed at 10~kHz, where the transitions are not severely limited by the filtering bandwidth (see Appendix B for eye diagram data up to 1~MHz). 


Fig.~\ref{width} shows the jitter observed in the two measurement setups. One major feature of the jitter data is the time of the start of the distribution (see arrows in Fig.~\ref{width}). This shows an input-to-output waveform signal time of approximately 2.0~$\mu s$, including electrical delay, converter delays, and fluxon travel time. In both setups, this time is dominated by an electrical delay from the ~1 MHz filters. 

The CFR setup exhibits a much narrower jitter distribution. Most distribution data fits reasonably well to a Gaussian distribution (see caption for details). The Gaussian width for the CFR setup is approximately 6 times narrower than in the DP setup, indicating significantly lower jitter. Both setups have the same nominal filtering and attenuation. However, the CFR setup has 13~dB of input attenuation at low temperatures, whereas all of the attenuation is at room temperature in the DP setup. Having some attenuators at low temperatures allows lower noise at the chip than having all the attenuators at room temperatures. Thus, it seems that noise through the attenuators may limit the jitter in the DP setup.

\section{Discussion} 
\begin{figure}[htbp]
\centerline{\includegraphics[width=0.8\columnwidth]{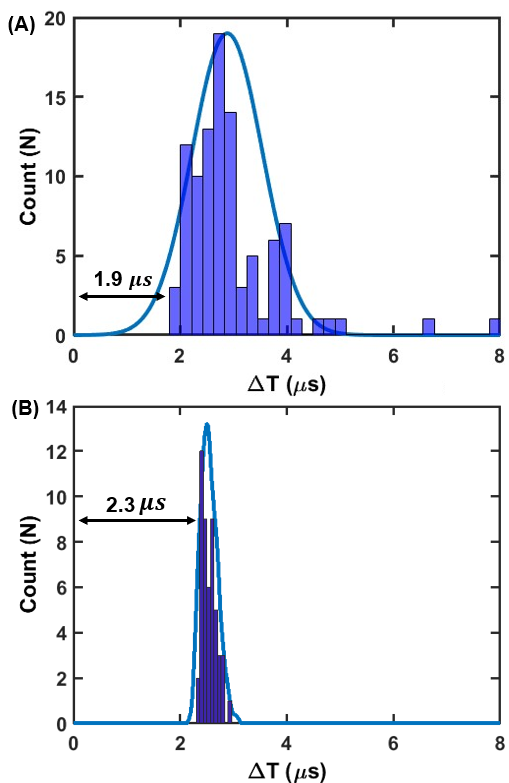}}
\caption{(A, B) Time delay distribution (jitter) generated from input and output waveforms, indicating the launch and measured detection times (see main text). Jitter is analyzed for an operating frequency of 10~kHz. In both setups, there is approximately 2~$\mu$s delay caused by the electrical delay of the input and output filters. Gaussian fits are made to give an approximate quantitative understanding. (A) Time delay distribution in the DP setup. A wide distribution suggests that the jitter (distribution width) is sensitive to the environment (see main text). Here we excluded 3 data points to the Gaussian fit between 6 and 10 $\mu s$ because they were outliers. The fit yielded a standard deviation of $\sigma=$ 0.66~$\mu$s. (B) Time delay distribution in the CFR setup. No points were excluded from the fit, and the fit yielded a standard deviation of $\sigma=$ 0.14~$\mu$s. The relatively narrow distribution suggests an improved measurement environment for the LJJ and detector compared to the DP setup. This narrow distribution shows how jitter can be useful for analyzing noise in the two experimental setups.
}
\label{width}
\end{figure}

The lower jitter from the CFR setup may be attributed to the lower noise at the fluxon detection circuit rather than the jitter from bulk LJJ transmission. We compare the observed delay time and jitter with the theoretical expectation for fluxon propagation in an LJJ. In an LJJ with negligible damping, assuming a speed of $v=0.78 c$, the fluxon takes a time $t_{\text{LJJ}} = 0.20 \text{ns}$  to traverse the LJJ of length $79 a$=1300 $\mu \text{m}$. This is orders of magnitude smaller than the filter-induced electrical delay and smaller than the optimized measured jitter in Fig.~\ref{width}. In an LJJ with finite damping rate $\alpha$ (arising e.g. due to dielectric losses related to the loss tangent of the JJs), the fluxon motion is ballistic within the time $t \ll 1/(2\pi \nu_J \alpha),$ but it turns diffusive in the time $t \gg 1/(2\pi \nu_J \alpha )$. Here $\nu_J= \sqrt{I_c/(2\pi \Phi_0 C_J)}=44$~GHz is the Josephson frequency and $C_J=300$~fF. Thus, for $t_{\text{LJJ}} = 0.20\, \text{ns}$, one may define a maximum damping rate $\alpha_{\text{max}} = 1/(2\pi \nu_J  t_{\text{LJJ}}) = 0.018$ below which the fluxon motion through our LJJ is likely ballistic. This is much higher than expected from dielectric losses in the JJs of the LJJ. If, on the other hand, a damping rate of $\alpha = 2\cdot 10^{-3}$ from dielectric losses is applied here \cite{khalil2013evidence}, ballistic fluxon motion is expected over a time scale $\ll 1/(2\pi \nu_J \alpha) = 1.8 \text{ns}$, which is much larger than $t_{\text{LJJ}}$ estimated above. 

Using this value of $\alpha$ and the dunk probe temperature of $4.2 \text{K}$, we can estimate the LJJ-jitter using a standard model for the fluxon propagation through an LJJ under thermal noise \cite{FedorovETAL2007}). For $t_{\text{LJJ}} = 0.20 \text{ns}$, the corresponding jitter is $\sigma_{\text{LJJ}} = 1.8 \text{ps}$ -- an entirely negligible value compared with the observed jitter in Fig.~\ref{width}.  Even if the damping rate is much larger (such as the maximum damping rate $\alpha_{\text{max}}$), our estimates for the LJJ transmission time and LJJ jitter are much smaller than the observed delay and jitter. In our measurements, the jitter is dominated by the detector circuit noise rather than the bulk LJJ transmission. Evidence of this is provided by a strong dependence of the measured jitter on the bias currents to the output converter, as shown in App.~\ref{app:jitter_biasdependence}.

The relatively small jitter in the CFR setup, with other knowledge from the circuit, indicates that there should be good ballistic motion of fluxons through the LJJ in the CFR setup, even though we have not experimentally verified it. If the minimum jitter in the detector were greatly reduced, we might verify ballistic dynamics through jitter $\sigma_{\text{LJJ}}$ from measurements of different LJJ lengths. According to theory, LJJ jitter $\sigma_{\text{LJJ}}$ scales with the transmission time $t_{\text{LJJ}}$ according to $\sigma_{\text{LJJ}} \propto t_{\text{LJJ}}^{3/2}$ and $\sigma_{\text{LJJ}} \propto t_{\text{LJJ}}^{1/2}$, for ballistic and diffusive dynamics, respectively. 
 
According to our measurements of inductance, our engineered LJJs have a rest energy of $E_{FL} (v=0)$ = $8 I_C \Phi_0 \lambda_J/2\pi a \approx 47$~zJ, where $\Phi_0$ is the flux quantum. If the fluxons were carrying 20\% of their total energy as kinetic energy \cite{wustmann2020reversible} (correspond to a fluxon velocity of $v=0.6c$), the total fluxon energy would be $E_{FL} (v=0.6c)$ = $10 I_C \Phi_0 \lambda_J/2\pi a\approx 59$~zJ. When simulating the circuit of Fig.~\ref{circuit} (A), we find that the fluxons were instead launched from the DC/SFQ converter at a speed of $v\approx 0.78c$, with the kinetic energy to rest energy ratio of $1/\sqrt{(1-0.78^2)}-1=60\%$. The energy loss for a stopped fluxon is then $60\%\times47$~zJ$\approx28$~zJ. The detection of fluxons with finite momentum (as performed by our detector), implies an energy loss much smaller than 28~zJ.

According to calculations on discrete LJJs, our fluxons at a speed of $v=0.78c$ should experience a fractional energy loss of $10^{-3}$ within a period of $\nu_J^{-1}$.  At this energy loss rate, we calculate a 3$\%$ change in energy corresponding to a travel distance of 2060~$\mu$m over a time of 682~ps. Our 1300 $\mu \text{m}$-long LJJ is shorter. As a result, we obtain a fluxon energy loss on the order of 1~zJ. 

In the future, we may lower discreteness to $L/L_J=1/7$ and lower launch velocity to $v\approx0.6c$. These two parameter changes will reduce the calculated plasma wave loss by approximately three orders of magnitude \cite{wustmann2020reversible}.

\section{CONCLUSION}

We detected low energy fluxons made using engineered LJJs, and studied the influence of their environments in two measurement setups. The fluxons are analogs to particles due to the nature of the fluxon in the sine-Gordon equation. In our discrete LJJ, they cause the coordinated evolution of JJs. At the beginning of the study, we estimated the discreteness of the LJJ from the ratio of the nominal Josephson junction inductance to the measured cell rail inductance ratio, $L_J/L$. In this study, this is slightly smaller than the intended value of 7. The preamplifier and other filtering in our setup limited the maximum fluxon measurement rate to $\sim$1~MHz. However, the fluxons travel fast and can in principle be spaced in time within an order of magnitude of the inverse of the Josephson frequency, which is $\nu_J$=44~GHz in this experiment.  

We measured fluxon arrival events after launching at a speed of 0.78c. Based on inductance measurements, we found that the rest energy of the fluxons is approximately $47~$zJ. Based on fluxon arrival measurements and calculations, our CFR setup allows ballistic fluxon motion, but the jitter of the fluxon motion is negligible in our measurement. Finally, the calculated energy loss from plasma waves is on the order of $1$ zJ.

The JJs in the detector have a small critical current of only 20 $\mu$A, which are susceptible to electrical noise from thermal noise or other noise brought from outside the cryogenic dewar. The CFR setup exhibits lower noise than the other setup. This is likely because the CFR setup provides a significantly improved environment for the detector, based on the measured sensitivity of the detector bias. The fluxons may be negligibly affected by the environment in our CFR setup, and are likely ballistically transferred. This study investigated jitter from low-energy fluxons. It is hopeful that the LJJ type, which is the engineered discrete LJJ, advances future tests in reversible logic. For RFL, engineered LJJs will provide the necessary nonlinearity for the required logic gate dynamics.

\section*{Acknowledgment}

The authors thank R. M. Lewis and M. P. Frank for providing a reference converter chip for testing our measurement setup, which we measured before tests of our chips containing LJJs. The authors also acknowledge scientific interactions with I. Vernik, A. F. Kirichenko, and D. Yohannes from SEEQC foundry services (www.seeqc.com). Furthermore, we thank C. Richardson, B. Palmer, and B. Butera from LPS for their helpful and collegial scientific interactions.

\appendices
\section{SQUID Measurements}

To test the inductance of a structure similar to the rail on our LJJs, we utilized an SQUID with electrodes to directly inject flux bias current into a metal strip in a DC SQUID. The inductance was determined by observing the magnetic field response of the SQUID to the injected flux bias current $I_{fl}$. When a current $I_{fl}$ is directly injected into the shared superconducting strip, each additional magnetic flux quantum coupled to the SQUID equals the product of $\Delta I_{fl}$ and the self-inductance of the shared strip, where the $\Delta I_{fl}$ is the period observed in the SQUID voltage modulation. Therefore, the self-inductance is given by the flux quantum $\Phi_0$ divided by $\Delta I_{fl}$.

\begin{figure}[htbp]
\centerline{\includegraphics[width=1\columnwidth]{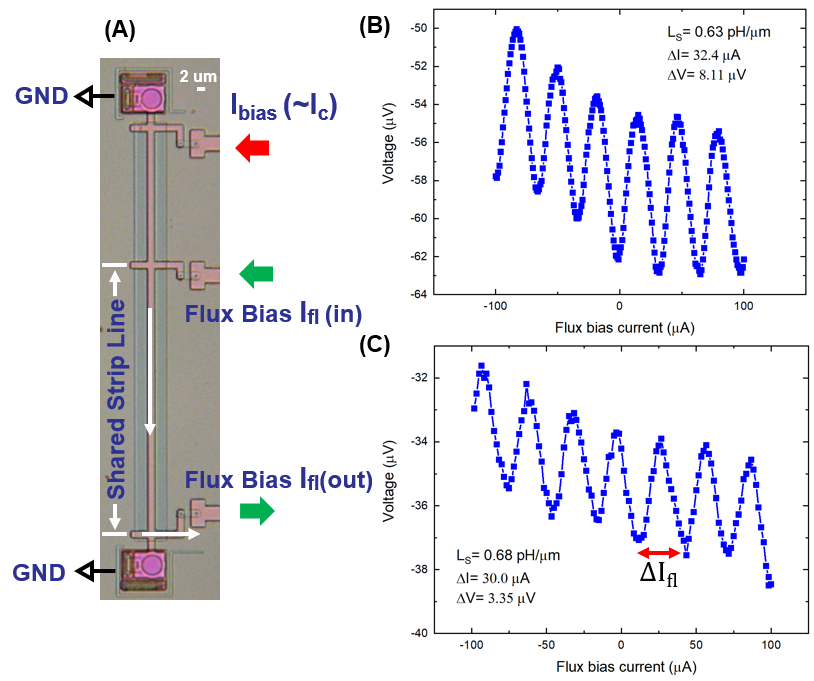}}
\caption{SQUID test data taken with the DP setup. (A) Optical photograph of Nb SQUID device used in inductance tests. Under the strip wiring of the SQUID, there is no ground plane. (B) and (C) show the voltage modulation of voltage-injected flux bias current of two nominally identical SQUIDs from two chips on the fabricated wafer. }
\label{SQUID}
\end{figure}

In Fig.~\ref{SQUID} (A), we present photographs of the test SQUID. It was cooled in the DP setup and statically biased with a constant current just above the critical current. Fig.~\ref{SQUID} (B) and (C) show examples of voltage responses to the flux bias current for two identically designed SQUIDs. The self-inductance per $\mu m$ was extracted using the formula $L_s=(\Phi_0/\Delta I_{fl})/l$, where $l$ represents the length of the shared strip. From these tests, we obtained an average inductance per $\mu m$ at 0.654~pH/$\mu m$, with a standard deviation of 3.8$\%$.
 
\section{Eye Diagrams}

\begin{figure}[htbp]
\centerline{\includegraphics[width=1\columnwidth]{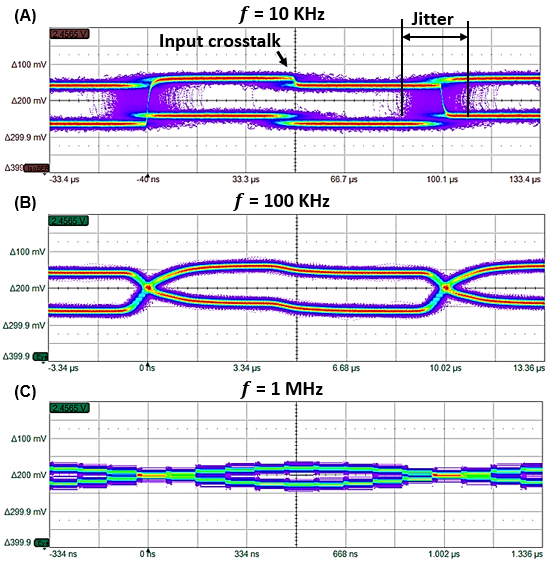}}
\caption{Eye diagram data captured on an oscilloscope in the CFR setup. The readout distinguishability of the SFQ/DC converter lowers as the frequency increases towards 1~MHz. The jitter of switching at this frequency is explained in the main text.}
\label{eyediagram}
\end{figure}

An oscilloscope was used to analyze jitter and eye diagrams. The eye diagram is a visual representation of the signal, with repeated cycles overlaid to show multiple time intervals of the signal. This can show the causes of the bit error rate. Fig. \ref{eyediagram} displays the eye diagram of the output signal under the CFR setup, obtained from an oscilloscope with input frequencies of 10~kHz, 100~kHz, and 1~MHz. The number of unit intervals overlaid on the eye diagram for analysis is 3.8~k, 19.6~k, and 73.3~k. The signal undergoes attenuation by a 3~dB attenuator and is amplified by 2000 by a preamplifier, resulting in an approximate overall amplification factor of 1678. Cross-talk is observed between the input and output waveforms, as a small break in the middle of the 10~kHz output waveform.

\section{Output amplitude}

\begin{figure}[htbp]
\centerline{\includegraphics[width=1\columnwidth]{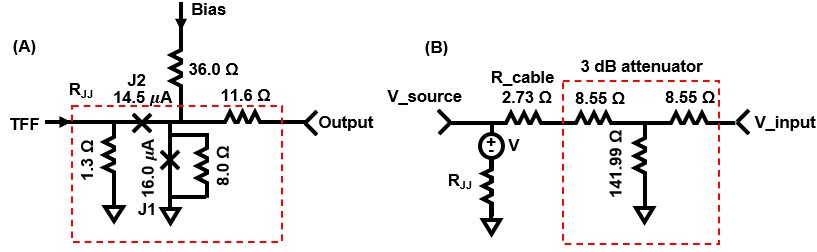}}
\caption{(A,B) Output circuit schematics. (A) The schematic of the detector output. The junctions utilized at the detector output have a critical current of approximately 15~$\mu$A. This is approximately ~25$\%$ smaller than the first JJ of the detector. (B) A connection circuit between low temperature and room temperature. $R_\text{cable}$ includes the chip-probe contact resistance and the cable resistance. $R_\text{JJ}$ corresponds to the detector impedance, which functions like an internal impedance of a voltage source ($V_\text{source}$). $V_\text{input}$ represents the input voltage of the preamplifier, whose input impedance is high.}
\label{output}
\end{figure}

Fig. \ref{output} shows the circuit schematic between the LJJ detector output and the room temperature electrical components. The LJJ detector output serves as the voltage source. A low-noise voltage preamplifier is set to a gain of 2000 and an input impedance of 100~M$\Omega$ is present. The detector output $V_\text{source}$ is divided down at the preamplifier input $V_\text{input}$ according to: $V_\text{input}$=$V_\text{source}\times\dfrac{141.99\Omega}{(141.99+8.55+2.73+R_\text{JJ})\Omega}=0.84V_\text{source}$. $R_\text{JJ}$ represents the internal impedance of the voltage source, which is equivalent to the impedance of the detector's output resistance, estimated to be approximately 16~$\Omega$.  $V_\text{input}$ equals the measured amplified value divided by the gain. From this, we extract the proper scaling factor between the oscilloscope's voltage and the detector's output voltage.

\section{The influence of bias currents on jitter}\label{app:jitter_biasdependence}

\begin{figure}[htbp]
\centerline{\includegraphics[width=0.7\columnwidth]{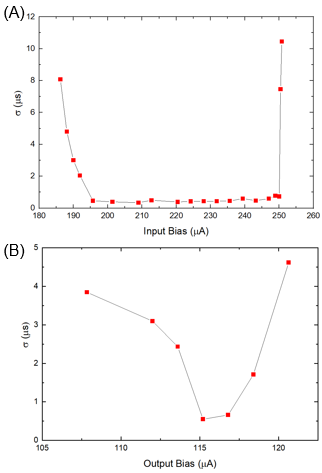}}
\caption{The extracted jitter for low-energy fluxon transmission in the dunk-probe setup. (A) The jitter as a function of the input-bias current. (B) The jitter as a function of output-bias current. }
\label{biasjitter}
\end{figure}

Fig. \ref{biasjitter} shows the effect of the applied bias currents on the extracted jitter. In general, under- or over-biasing leads to increased jitter. We define input bias as the single bias sent to the launcher, composed of the DC/SFQ converter and the JTL stage. Incorrect input bias to the launcher may create an inconsistent delay in launches. This may increase jitter from the lowest value (see panel A). A low-jitter, seen at $\sigma ~ 0.66 \mu s$, is observed over an input bias range of $222 \mu A \pm 12\% $. On the other hand, the output bias goes to the detector and affects the detection events. The detector has a small critical current that may cause noise in the fluxon velocity as it enters the detector. As shown in panel B), the jitter quickly increases above the minimum ($\sigma = 0.66 \mu s$) for any change in output bias. At bias values of 115 $\mu$A to 118 $\mu$A, the jitter is 3 times smaller than other points. The great sensitivity here is possibly due to the much smaller critical current of the JJs used in the detector relative to the DC/SFQ launcher. Specifically, the first JJ in the detector has a 20 $\mu A$ critical current and all JJs in the launcher are approximately 100 $\mu A$. This gives evidence that thermal noise increases jitter in the detector. This is not surprising because a small JJ is necessary in our experiment to measure the arrival of a low-energy fluxon. Note that $\lambda_J$ is only 2.4 unit cells in our LJJ, and each JJ of the LJJ has only 7.5 $\mu A$ critical current, indicating our low energy scale. If these measurements are redesigned in the future, the detector should be improved upon to allow lower jitter measurements.   


\end{document}